\documentclass[aps,prd,twocolumn,superscriptaddress,showkeys,showpacs]{revtex4}
\usepackage{mathptmx}
\usepackage[utf8]{inputenc}
\usepackage[T1]{fontenc}
\usepackage{slashed}
\usepackage{times}
\usepackage{amssymb,amsfonts,amsmath,amsthm}
\usepackage{dsfont,bbm}
\usepackage{dcolumn}
\usepackage{epsf}
\usepackage{graphicx}
\usepackage[caption=false]{subfig}
\usepackage{dsfont}
\usepackage{slashed}
\usepackage[active]{srcltx}
\usepackage[usenames]{color}

\begin{document}

\title{The contour gauge in use: telling untold}

\author{I.V.~Anikin}
\email{anikin@theor.jinr.ru, Igor.Anikin@ncbj.gov.pl}
\affiliation{Bogoliubov Laboratory of Theoretical Physics, JINR,
             141980 Dubna, Russia}
\affiliation{National Centre for Nuclear Research (NCBJ),
            00-681 Warsaw, Poland}

\begin{abstract}
In Quantum Field Theory, we discuss the main features of the (non-local) contour gauge
which extends the local axial-type gauge used in most approaches.
Based on the gluon geometry,
we demonstrate that the contour gauge does not suffer from the residual gauge.
We discuss the useful correspondence between the contour gauge conception and the
Hamiltonian (Lagrangian) formalism. Having compared
the local and non-local gauges, we again advocate the advantage of the contour gauge use.
\end{abstract}
\date{\today}
\maketitle

\section{Introduction}

Since any gauge theories can be considered as the systems with dynamical constraints,
it is difficult to overestimate the role of gauge conditions
which are nothing but the additional constraints appeared in the Hamiltonian formalism.
As well-known, the gauge theory can be consistently quantized if it is possible to uniquely resolve all constraint conditions
eliminating the unphysical degrees of freedom. However, in the Hamiltonian (Lagrangian) approach,
the constraint conditions with respect to the generalized momenta and coordinates
have, as a rule, very nontrivial forms and, hence, finding unique solutions of these equations is not a simple task.
It sometimes becomes even impossible in some cases due to the presence of the Gribov ambiguities.

Fortunately, as it is known for many years, it turns out that the infinite group orbit volume, which leads to the problem with quantization,
can be factorized out to the insubstantial normalization factor thanks to the gauge invariance property
of the corresponding Lagrangian (or Hamiltonian) of a theory \cite{Faddeev:1980be}.
The mentioned infinite volume that has been included in the
factorized prefactor of the functional integration
is now not the principle obstruction for the quantization of the gauge theory.

Nowadays, only the local type of gauges can traditionally be called as the most popular gauges
used for the practical applications.
At the same time, some of useful local gauges, for example the axial-type of gauges, 
might suffer from the
residual gauge freedom which, in its turn, can ultimately give rise to the spurious singularity.
In contrast to the local (axial) gauge condition, the use of contour gauge, as a class of non-local gauges, allows us indeed
to fix completely the gauge freedom in the most simply form without an additional assumptions. 
In this case, there is no the residual gauge freedom unless the boundary conditions for gluons
relate to the non-trivial topology (see also \cite{Anikin:2021oht}).  

Notice that this complete gauge fixing is provided by the construction of contour gauge from the very beginning.
Indeed, within the contour gauge conception
the gauge orbit representative should be first fixed and, then, a certain local gauge condition
which is correlated with the given (gauge) orbit representative should be searched.
In some sense, the opposite logic of the gauge fixing compared to the use of usual local gauges takes place.

Among the modern phenomenology theorists,
there is such a prejudice that the study of any non-local gauge conditions in QCD  
is not attractive because specific contour-gauge techniques adopted for the QCD phenomenology 
are very complicated and the efforts to study the differential geometry details cannot be requited.
In the present short paper, we attempt to break this superficial and wrong impression.
Namely, the differential geometry technique and the interpretation of gluons   
as a connection on the principle fiber bundle are merely needed {\it (a)} for the demonstration 
that with the help of the contour gauge
the gauge freedom can be uniquely fixed upto the residual freedom;  
{\it (b)} for a proof of the fact that in the local gauge, given by $A^+=0$,
two different representations of transverse gluon field through the strength tensor,
related to the integrations from $x$ to $+\infty$ and from $-\infty$ to $x$,
are actually not equivalent ones. We stress that without the contour gauge which is more general 
in comparison with any axial local gauges, it is not possible 
to distinguish these two different representations. At the same time, 
the assumption of equivalence leads to many problems, for example, with the gauge invariance of 
Drell-Yan-like hadron tensors.  

Since the interest to the details of contour gauge applications
increases in the phenomenological community, 
it is worth to make public the important subtleties based on the mathematical   
technique adjusted to the physical language.  
Moreover, we focus here on the important explanations which are being remained untold ones in the preceding publications.
On the other hand, since in the recent literature 
one can still find a wrong representation of the transverse gluon field through the strength tensor
considered in the local axial gauge, we treat this fact as a lack of explanation of what means the contour gauge and of
how to use it. All these underlie our motivation of the present paper.

In the first part of the paper, in order to clarify the main features of contour gauge as a gauge without the residual gauge freedom,
we shortly remind  the elements of Hamiltonian and Lagrangian formalisms of quantization. 
Actually, the Hamiltonian formalism might be considered as something useless from the point of view of 
phenomenological applications. But, on the other hand, the Hamiltonian formalism is 
a more convenient frame to understand the contour gauge in the context that how
the contour gauge condition defines the manifold surface crossing over a group orbit of the fiber uniquely.
Meanwhile, there are no doubts that the Lagrangian formalism is assumed to be better designed for the practical applications.
We combine both formalisms depending on the current goals.

The attempt to present an alternative explanations of the contour gauge conception has been done in the second part of the paper.
Apart, we also implement a comparative analysis of the local and non-local gauges of axial type.

\section{The Lagrangian and Hamiltonian systems with the dynamical constraints}
\label{Sec:II}

As well-known, the local axial gauge, $A^+=0$, suffers fr\-om the residual gauge freedom
demanding the additional requirements to fix it.
In the most cases, the formulation of additional requirements    
is not a trivial task within the Lagrangian system.
Indeed, if one demands simultaneously $A^+=A^-=0$, the maximal gauge fixing is
got no problem in a classical theory only.
However, in a quantum theory, the simultaneous conditions $A^+=0$ and $A^-=0$ 
as delta-function arguments in the corresponding 
functional integration (which lead to the effective Lagrangian with 
$1/\xi_1 (n\cdot A)^2$ and $1/\xi_2 (n^*\cdot A)^2$) 
result in the absence of the well-defined gauge propagator 
because the corresponding kinematical operator cannot be inverted.
In this connection, it is necessary to establish 
an alternative method of gauge fixing compared to the “classical” approaches with the effective Lagrangian
including both $(n\cdot A)^2$ and $(n^*\cdot A)^2$.
The contour gauge conception gives us such an alternative and effective method. 

To better understanding of subtleties related to the contour gauge, 
it is worth to remind the main stages of the Lagrangian (L-system) and Hamiltonian (H-system) approaches to the quantization of gauge fields
(see, for example, \cite{Konopleva:1981ew}).
In this section, we use the textbook language to show that the H-system is the most adequate approach 
in order to demonstrate the gauge-field quantization ``philosophy'' with the resolved additional constraints.  
However, due to the special role of a time-coordinate 
in H-system, the L-system is more suitable for a practical use.
In other words, the H-system is needed for a help to see how the contour gauge fixes uniquely the total gauge freedom,
but the main computation procedure has been formulated in terms of L-system. 

\subsection{H-system with dynamical constraints}

Let us consider the H-system, defined by $H(p_i, q_i)$, where the phase space $\Gamma$ has been formed by
the generalized momenta $p_i$ and coordinates $q_i$ and, in addition,
$2m$ constraints on $p_i$ and $q_i$ have been imposed.
Traditionally, these $2m$ constraints are denoted as $\varphi_a(p_i, q_i)$ and $\chi_a(p_i, q_i)$.

We suppose that the H-system we consider has an equivalence orbit which is nothing but the gauge group orbit in the gauge theory.
Since the phase space $\Gamma$ is overfilled by unphysical degrees, in the most ideal case, we have to resolve all the constraints.
The additional constraints $\chi_a(p_i, q_i)$ are necessary to fix uniquely the orbit representative in order to quantize the H-system.
After resolving all constraints, we deal with the quantized H-system where the physical phase space $\Gamma^*$ of dimension $2(n-m)$
is a subspace of the initial space $\Gamma$ of dimension $2n$ and is a fundament for H-system in terms of physical configurations,
$H^*(p_i^*, q^*_i)$.

Hence, in terms of the functional integration, the amplitude between the initial $| q_1^{in}, ..., q^{in}_n; t^{in} \rangle$ and the
final $\langle q_1^f, ..., q^f_n; t^f |$ states takes the following form (modulo the unimportant normalization factors)
\cite{Konopleva:1981ew,Faddeev:1980be}
\begin{eqnarray}
\label{Amp-1}
&&\langle q_1^f, ..., q^f_n; t^f | q_1^{in}, ..., q^{in}_n; t^{in} \rangle =
\nonumber\\
&& N\,
\int {\cal D} p_i(t)  {\cal D} q_i(t) \, \delta(\varphi_a) \delta(\chi_a)\, \text{det} \big\{ \varphi_a, \chi_a \big\} \times
\nonumber\\
&&\text{exp} \Big\{
i\int_{t_{in}}^{t_{f}} dt \big[ p_i \partial_0 q_i  - H(p_i, q_i) \big]
\Big\},
\end{eqnarray}
where $\{ ..\, , \, ..\}$ denotes the Poisson brackets.

As usual, the delta-function $\delta(\varphi_a)$ of Eqn.~(\ref{Amp-1}) can be presented through
the integration over the Lagrange factor $\lambda_a$ as
\begin{eqnarray}
\label{Lag-1}
\delta(\varphi_a)= \int (d\lambda_a) \,e^{i \lambda_a \, \varphi_a(p_i, q_i)}
\end{eqnarray}
giving the generalized Hamiltonian of system which reads
\begin{eqnarray}
\label{H-prime}
H^\prime (p_i, q_i) = H(p_i, q_i) + \sum_a \, \lambda_a \, \varphi_a(p_i, q_i).
\end{eqnarray}

If we now suppose that the constraint conditions (see the delta-function arguments)
have somehow been resolved,
the amplitude is given by the following functional integration
\begin{eqnarray}
\label{Amp-1-2}
&&\langle q_1^f, ..., q^f_n; t^f | q_1^{in}, ..., q^{in}_n; t^{in} \rangle =
 N\,
\int {\cal D} p^*_i(t)  {\cal D} q^*_i(t) \times
\nonumber\\
&&
\text{exp} \Big\{
i\int_{t_{in}}^{t_{f}} dt \big[ p^*_i \partial_0 q^*_i  - H^*(p^*_i, q^*_i) \big]
\Big\},
\end{eqnarray}
where only the physical generalized momenta and coordinates are forming the integration measure and the
Hamiltonian. In other words, the generating functional of (\ref{Amp-1-2}) corresponds to the H-system 
with the dynamical constraints which have been resolved and, therefore,
there is no a (gauge) freedom associated with the arbitrary Lagrange factor $\lambda_a$.
It would be an ideal situation which, as known, cannot be realized practically in the most of cases.
However, the contour gauge as a class of non-local gauges gives a possibility to realize practically 
the mentioned ideal situation because the contour gauge condition has a unique solution by construction, see below.  

\begin{figure}[t]
\centerline{\includegraphics[width=0.45\textwidth]{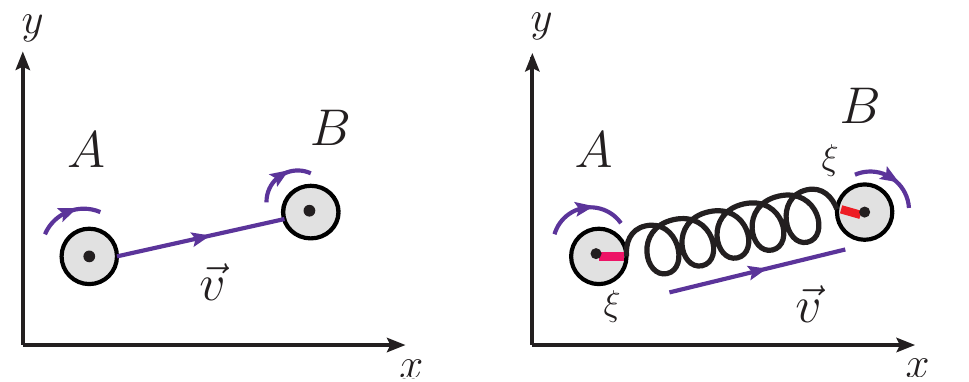}}
\caption{The H-system with unresolved and resolved constrains. The left panel represents
the system with the inner (gauge) symmetry; the right panel corresponds to
the system with the fixed (preferable) site $\xi$ on the ball surface.}
\label{Fig-S-1}
\end{figure}

It is instructive to illustrate the difference between the H-system with unresolved and resolved constraint
conditions with the help of a trivial mechanical example, see Fig.~\ref{Fig-S-1}.
Let us consider the homogenous ball which moves from the point $A$ to the point $B$.
The ball has a spherical symmetry under the rotation around the inertia center.
For the simplicity, we focus on the rotation of the
ball in some plane.
Since the ball surface is homogenous, we do not have a chance to observe
the rotation of the moving ball unless we mark some point on the surface.
The invisible ball rotation around its center of inertia corresponds, roughly speaking, to the inner (gauge)
transformations of the H-system and does not affect much the trajectory provided
the angle velocity is constant, see the left panel of Fig.~\ref{Fig-S-1}.
In this case, the inertia center plays a role of ``physical'' configurations of the H-system, while
the moving of different sites on the ball surface is invisible and relates to the ``unphysical'' configurations.

If we mark the site on the ball surface by a dash, we break down the
rotation symmetry in the meaning of that we choose the preferable site of the ball surface
and the ball rotation becomes visible.
Therefore, in this case, we can describe the moving dash together with the inertia center
as ``physical'' configurations of the H-system when the ball position varies from $A$ to $B$,
see the right panel of Fig.~\ref{Fig-S-1}.

In the context of the contour gauge conception, the marked dash on the ball corresponds to the 
certain gauge function $\theta$ of the photon (gluon) gauge transforms which has been fixed by the contour gauge condition.  
Moreover, in the above mentioned ideal case of resolved constraints, 
we would merely deal with the photon (or gluon) as a massless vector field
(transforming as a spinor of $2$-rank under the Lorentz group) which is described by the Hamiltonian
without the gauge transforms.

Notice that the Abelian and/or non-Abelian gauge theory can be treated as the H-system where
the amplitudes of the state transitions
take the forms which are similar to Eqn.~(\ref{Amp-1}) provided the following correspondences
between the canonical variables and the dynamical constraints (see below)
\begin{eqnarray}
\label{Rep-1}
\big( p_i;\, q_i  \,; t \big) &\Longrightarrow& \big( (E_0, E_i) ; \, (A_0, A_i) \,; x \big),
\nonumber\\
\delta(\varphi_a) &\Longrightarrow& \delta\left(  \partial_i E_i \right) \, \delta\left( E_0\right),
\nonumber\\
\delta(\chi_a) &\Longrightarrow& \delta\left(  \partial_iA_i \right) \, \delta\left( A_0\right),
\nonumber\\
\text{det} \big\{ \varphi_a, \chi_a \big\}  &\Longrightarrow& \Phi(A).
\end{eqnarray}

%
%
%

\subsection{L-system with dynamical constraints and its connection with H-system}

In the most typical gauge theories, resolving the additional (gauge) conditions is not a simple task.
The gauge conditions, leading to the equation system for the gauge function $\theta(x)$, would have an unique
solution $\theta_0$. In fact, it might be practically impossible.

To avoid the need for finding the unique solution $\theta_0$, Faddeev and Popov have proposed
the approach (FP-method) where
the infinite group orbit volume can be factorized out to the insubstantial normalization factor thanks
to the gauge invariance property
of the corresponding Lagrangian (or Hamiltonian) of theory  \cite{Faddeev:1980be}.

For the sake of simplicity, we are now going to dwell on $U(1)$ gauge theory.
The H-system can be readily stemmed from the Lagrangian formalism of
first order where the Lagrangian is defined as
\begin{eqnarray}
\label{Lag-1ord}
{\cal L}_{(1)}= -\frac{1}{2} \Big[
\partial_\mu A_\nu(x) - \partial_\nu A_\mu - \frac{1}{2}F_{\mu\nu}
\Big] F_{\mu\nu},
\end{eqnarray}
with $A_\mu$ and $F_{\mu\nu}$ being independent field configurations. The Lagrangian of Eqn.~(\ref{Lag-1ord})
can be written in $3$-dimensional form as
\begin{eqnarray}
\label{Lag-1ord-2}
{\cal L}_{(1)}= E_i \,\partial_0A_i + A_0 \,\partial_i E_i + \frac{1}{2} E^2
+ \frac{1}{4} B_i \, \epsilon_{i k \ell }\partial_k A_\ell + \frac{1}{2} B^2,
 \end{eqnarray}
where $E_i=F_{i 0}$ and $A_i$ imply the generalized momenta and coordinates, respectively;
$B_i=1/2\,\epsilon_{i j k} F_{jk}$ are the implicit variables in the integration measure, see below.
From the Euler-Lagrange equations we can get the equations which do not include $\partial_0$
and give the constraint conditions written in the forms of
\begin{eqnarray}
\label{Const-Cond}
&&E_0=0 \text{ -- the primary condition},
\nonumber\\
&&
\partial_i E_i =0 \text{ -- the secondary condition} .
\end{eqnarray}

As well-known, any gauge theory can be considered as the theory with the constraints which
are applied on the field configurations.
Moreover, the constraint conditions of Eqn.~(\ref{Const-Cond}) have to be completed by the additional (gauge) conditions,
for example
\begin{eqnarray}
\label{Gauge-Cond}
A_0=0 \,\,\, \text{and}\,\,\,   \partial_i A_i =0.
\end{eqnarray}
From the theoretical point of view, the full set of conditions defined by Eqns,~(\ref{Const-Cond}) and (\ref{Gauge-Cond})
should eliminate all unphysical degrees of freedom for the correct quantization of H- (or L-) system.

Using the FP method,  we adhere the functional integration approach to quantization of gauge fields and
start from the functional integration written for the L-system as \cite{Konopleva:1981ew}
\begin{eqnarray}
\label{FI-L-1}
\int {\cal D}\theta(x) \, \int {\cal D}A_\mu \,\delta\left( F[A] \right) \Phi(A)\, e^{i S[A]},
\end{eqnarray}
where the infinite group orbit volume given by ${\cal D}\theta(x)$
has been factorized out in the integration measure
due to the gauge invariant action $S[A]$ and functional $\Phi(A)$.
Hence, the exact magnitude of the group volume prefactor, {\it i.e.}
\begin{eqnarray}
\label{G-orbit}
v_g\equiv \int {\cal D}\theta(x) =\int \prod_x [d\theta(x)]= \big\{ \infty, \,0\big\},
\end{eqnarray}
is irrelevant because
the prefactor should be cancelled by the corresponding normalization of Green functions.

In Eqn.~(\ref{G-orbit}), the infinite group volume corresponds to the standard case of unresolved gauge conditions,
while the zero group volume appears in the case of resolved gauge conditions. 
The latter is a principal observation of this section.
Indeed,
the integration measure  $[d\theta(x)]$ can be defined on the group manifold as an invariant measure,
\begin{eqnarray}
\label{G-inv-m}
\int [d\theta] f(\theta) \sim \sum_{\theta\in G} f(\theta).
\end{eqnarray}
$\prod_x [d\theta(x)]$ denotes the product of the invariant measures
defined on the structure group $G$ of fiber over each point of the Minkowski space.
If the gauge function $\theta$ is not fixed, we have the infinite integration over
the group invariant measure,
otherwise (that is, if the gauge function $\theta$ is somehow fixed) the integration is equal zero for
each point of the Minkowski space but the value of fixed $\theta_i$ can vary from one point to another.

Eqn.~(\ref{FI-L-1}) can be identically rewritten as
\begin{eqnarray}
\label{FI-L-2}
\int {\cal D}A_\mu {\cal D} F_{\mu\nu} \,\delta\left( F[A] \right) \Phi(A)\, e^{i S[A, F]},
\end{eqnarray}
where the functional of action $S[A, F]$ is given the Lagrangian of Eqn.~(\ref{Lag-1ord}).
Focusing on the three-dimensional forms, we obtain that the functional integration is given by
\begin{eqnarray}
\label{FI-L-3}
\int {\cal D}A_i  {\cal D}A_0 \,
 {\cal D} B_i   {\cal D} E_i \,\delta\left( F[A] \right) \Phi(A)\, e^{i S[A, E, B]},
\end{eqnarray}
where $S[A, E, B]$ is now defined by Eqn.~(\ref{Lag-1ord-2}).
Integrations over $B_i$ and $A_0$ in Eqn.~(\ref{FI-L-3}) lead to the following functional integral:
\begin{eqnarray}
\label{FI-L-4}
&&
\int {\cal D}A_i \,
{\cal D} E_i \, \delta\left( F[A] \right) \Phi(A) \delta\left(  \partial_i E_i \right)
\times
\nonumber\\
&&
\text{exp}\Big\{
i \int dz \Big[
E_i \,\partial_0A_i + H(E_i, A_i)
\Big]
\Big\},
\end{eqnarray}
where the Hamiltonian is given by
\begin{eqnarray}
\label{H-1}
H(E_i, A_i) = \frac{1}{2} E^2
- \frac{1}{2} \left( \epsilon_{i j k} \partial_j A_k\right)^2
\end{eqnarray}
and
the gauge condition can be written as $F(A)= \partial_i A_i$.
On the other hand, Eqn.~(\ref{FI-L-4}) can be presented in the equivalent form as
\begin{eqnarray}
\label{FI-L-5}
&&
\int {\cal D}A_i  {\cal D}A_0\,
{\cal D} E_i {\cal D} E_0\, \delta\left(  \partial_iA_i \right) \Phi(A) \delta\left(  \partial_i E_i \right)
\delta\left( A_0\right)\delta\left( E_0\right)
\times
\nonumber\\
&&
\text{exp}\Big\{
i \int dz \Big[
E_i \,\partial_0 A_i + E_0\partial_0 A_0 + H(E_i, A_i; E_0, A_0)
\Big]
\Big\},
\end{eqnarray}
where the primary and secondary constraints together with the gauge conditions of Eqn.~(\ref{Gauge-Cond})
have been explicitly shown in the functional integrand.
This representation of H-system resembles the functional integration presented by Eqn.~(\ref{Amp-1}).

Notice that, in H-system, it is not a problem to write all constraints through the delta functions because 
we have no needs to invert the kinematical-like operator as it would be necessary in L-system forming the Feynman rules.
At the same time, the H-system approach is not convenient for the practical computation in QFT.

In conclusion, the basic reason for writing this section in the textbook language  
is the following: the section is preparing a reader for the main features of contour gauge uses.
Indeed, we have reminded the differences between the H-systems with and without resolved additional 
dynamical constraints presenting the mechanical illustration in Fig.~(\ref{Fig-S-1}).
As well known, within the FP-method of L- and H-systems there is no need to find a unique solution 
of  the system of the additional (gauge) conditions with respect to the gauge function $\theta(x)$.
Thanks for the gauge invariance of Lagrangian (Hamiltonian) and functional $\Phi(A)$,
the infinite group orbit volume defined by $[d\theta]$ has been factorized out in the corresponding
functional integration measure giving a possibility to quantize the gauge theory
modulo the residual gauge problems.
However, as mentioned,
if we resolve the additional (gauge) condition with respect to $\theta$, the group representative on each of orbit
can be fixed uniquely and the factorized group integral defined through the Riemann summations should be equal to zero.
Since, the group volume has been cancelled by the corresponding normalization, 
it does not mean that the functional integration of Eqn.~(\ref{FI-L-1})
disappears. This case takes place if we use the contour gauge conception.

\section{The contour gauge conception}
\label{Sec:III}

In this section, we concentrate on the description of the contour gauge conception
which implies that the corresponding gauge condition can be formally resolved in order to find
the unique gauge orbit representative.
A few decades ago, the contour gauge had intensively studied
due to the fact
that the quantum gauge theory should not suffer from the Gribov ambiguities (see, for example,
\cite{ContourG1, ContourG2,Shevchenko:1998uw}).

It states that the gauge function can be completely fix (in the H-system, see Eqn.~(\ref{Amp-1}))  or
the unphysical gluons can be eliminated (in the L-system, see Eqn.~(\ref{FI-L-1}))
if we demand the path dependent functional (Wilson path functional)
to be equal to unity, {\it i.e.}
\begin{eqnarray}
\label{CG-1}
{\bf g}(x | A)\Big |_{P(x_0,x)}
\equiv\mathbb{P}\text{exp} \Big\{ ig\int_{P(x_0,x)} d\omega_\mu A_\mu(\omega)\Big\}={\bf 1}
\end{eqnarray}
where the path $P(x_0,x)$ between the points $x_0$ and $x$ is fixed.
The axial (light-cone) gauge, $A^+=0$, is a particular case of the non-local contour gauge
determined by Eqn.~(\ref{CG-1}) provided the fixed path is given by the straightforward line
connecting $\pm\infty$ with $x$.
By construction, the contour gauge does not possess the residual gauge freedom in the finite space,
{\it i.e.} in what follows the boundary gluon configurations have assumed to be equal to zero
(see \cite{Anikin:2021oht} where it is shown that the possible residual gauge
can be located at the corresponding boundary).
It also gives, from the technical point of view, the simplest way to fix the gauge function completely.

The contour gauge conception is closely related to the path group formalism \cite{Mensky:2012iy,Mensky-book}
and can be traced from the Mandelstam approach \cite{Mandelstam:1962mi}.
The geometrical interpretation of gluon fields as a connection on the
principle fiber bundle ${\cal P}({\cal X}, \pi \,|\, G)$ is the underlying basis
for the use of non-local gauges.
In this connection, let us define two directions: one direction is determined in the base ${\cal X}$
as the tangent vector of a curve
going through the point $x\in {\cal X}$; the other
direction is defined in the fiber and can be uniquely determined as the
tangent subspace related to the parallel transport \cite{Mensky:2012iy,Mensky-book}.
These two directions allow us to introduce the horizontal vector defined as
\begin{eqnarray}
\label{H-vector}
H_\mu = \frac{\partial}{\partial x^\mu} -ig A^a_\mu(x) \, D^{a},
\quad D^{a}= {\bf D}^a \cdot {\bf g} \frac{\partial}{\partial {\bf g}},
\end{eqnarray}
where $D^a$ denotes the corresponding shift generator along the group fiber written in the differential form
and, together with the vector coefficients (connection of the principle fiber bundle)
$A^a_\mu(x)$, defines the algebraic vertical (tangent) vector field on the fiber \cite{Mensky:2012iy,Mensky-book}.
The horizontal vector $H_\mu$ is invariant under the structure group $G$ acting on the given representation
of the fiber by construction.

In ${\cal P}({\cal X}, \pi \,|\, G)$, the functional
${\bf g}(x | A)$ of Eqn.~(\ref{CG-1})
is a solution of
the parallel transport equation given by
\begin{eqnarray}
\label{H-vector-2}
\frac{d x_{\mu}(s)}{d s} H_\mu(A) {\bf g}(x(s)| A) = 0,
\end{eqnarray}
where the fiber point  $p(s)=\left( x(s), {\bf g}(x(s)) \right)$
with the curve $x(s)\in {\cal X}$ parametrized by $s$.
Eqn.~(\ref{H-vector-2}) being an algebraic differential equation takes place even if  ${\bf g}$ is fixed on the group
because, in this case, $D^a {\bf g}=0$ while $A^a_\mu(x)\not = 0$.
In this connection, the condition presented by Eqn.~(\ref{CG-1}) implies that the full curve-linear integration
goes to zero rather than the integrand itself.

\begin{figure}[t]
\centerline{\includegraphics[width=0.45\textwidth]{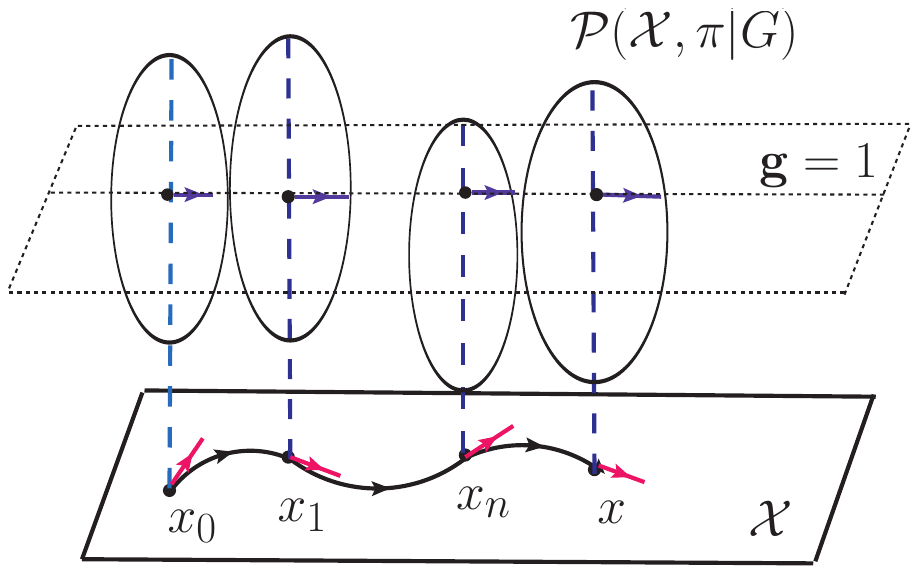}}
\vspace{-6.5cm}
\caption{The contour gauge: the plane of ${\bf g}=1$ in the principle fiber bundle ${\cal P}({\cal X}, \pi | G)$. }
\label{Fig-S-2}
\end{figure}

It can be shown \cite{Mensky:2012iy, Mensky-book} that every of points belonging to the fiber bundle, ${\cal P}({\cal X}, \pi \,|\, G)$,
has one and only one horizontal vector corresponding to the given tangent vector at $x\in {\cal X}$.
We remind that the tangent vector at the point $x$ is uniquely determined by the
given path passing through $x$.
In the frame of H-system based on the geometry of gluons
the condition of Eqn.~(\ref{CG-1}) corresponds to the determining of the surface
on ${\cal P}({\cal X}, \pi \,|\, G)$ that is parallel to the base plane with the path.
Moreover, it singles out the identity element, ${\bf g}=1$, in every fibers of ${\cal P}({\cal X}, \pi \,|\, G)$,
see Fig.~\ref{Fig-S-2}.
This choice can be traced to the Lagrange factor $\lambda_a$ which is formally
fixed in H-system, see Sec.~\ref{Sec:II}.
Roughly speaking, once the group (any) element of fiber is fixed, we deal with the Lagrange factor $\lambda_a$
of H-system which is also uniquely fixed.
On the other hand, if we fix the group element of fiber we fix the function theta of gauge transforms as well.
In this sense, we do not have the local gauge transforms (or the gauge freedom) anymore.
It resembles a bit the case of spontaneous breaking of symmetry.

Notice that the path dependent functional given by the {\it l.h.s.} of (\ref{CG-1}) can be also gauge transformed as
\begin{eqnarray}
\label{G-tr-wl}
&&\mathbb{P}\text{exp} \Big\{ ig\int_{P(x_0,x)} d\omega_\mu A^\theta_\mu(\omega)\Big\}=
\nonumber\\
&&
\omega(x)
\mathbb{P}\text{exp} \Big\{ ig\int_{P(x_0,x)} d\omega_\mu A_\mu(\omega)\Big\}
\omega^{-1}(x_0),
\end{eqnarray}
where the local gauge function defined by
\begin{eqnarray}
\label{U}
\omega(x)=e^{i\theta(x)}\equiv e^{iT^a \theta^a(x)}
\end{eqnarray}
with the corresponding generator $T^a$. From this, one can see if the Minkowski space has been realized as a loop space, 
the path dependent functional becomes invariant under the local gauge transforms.
In the general case of arbitrary paths, imposing the condition (\ref{CG-1}) on the {\it r.h.s.} of (\ref{G-tr-wl}),
we are able to get the different contour gauge ${\bf g}(x | A)=C$
with $C=\omega(x)\omega^{-1}(x_0)$, see (\ref{U}). From the theoretical point of view, this new contour gauge 
has the same status as the gauge of (\ref{CG-1}). 
It would correspond to the different plane ${\bf g}=C$ in Fig.~\ref{Fig-S-2} which also transects the principle fiber bundle
and, therefore, it generates the different contour gauge condition which, generally speaking, is related to the previous contour gauge by the
local transform. However, from the practical point of view, the contour gauge 
given by ${\bf g}(x | A)=C$ is not convenient to use for calculations because the representation of transverse gluon field 
through the strength tensor has a more complicated form compared to, for example, (\ref{Ag1}).  
Of course, the physical quantities are independent on the contour gauge choice.

Notice that since the gauge condition given by Eqn.~(\ref{CG-1}) selects only the
identity element of group $G$ on each fiber, it means that the gauge transforms have been reduced to the ``global''
gauge transforms. That is, the gauge function $\theta(x)$ becomes the coordinate independent and is fixed $\theta_0$.
This situation is identical to that one can see in Fig.~\ref{Fig-S-1} of Sec.~\ref{Sec:II}. 
Namely, the red dash on the ball surface corresponds to
one particular choice of the group element fixed by the given contour gauge. However, if we mark the line on the ball surface
by the other dash, it would mean that we choose the other group element fixed by the different contour gauge.
In both cases, we deal with the same description of H-system.

Since the functional ${\bf g}(x | A)$ depends on the whole path in ${\cal X}$,
the contour gauge refers to the non-local class of gauges and
generalises naturally the familiar local axial-type of gauges.
It is also worth
to notice that two different contour gauges can correspond to the same
local (axial) gauge with the different fixing of residual gauges \cite{Anikin:2010wz, Anikin:2015xka}.
If we would consider the local gauge without the connection with the non-local gauge,
the residual gauge freedom would require the extra conditions to fix the given freedom.
This statement reflects the fact that, in contrast to the local axial gauge,
the contour gauge does not possess the residual gauge freedom in the finite region of a space
where the boundary gluon fields are absent \cite{Anikin:2021oht}.
%
%
%
\section{Local and non-local gauge matching}
\label{Sec:IV}
Since the contour gauge as a non-local kind of gauges generalizes the standard local gauge of axial type,
it is worth to discuss shortly the correspondence between the local and non-local gauge transforms.

Let us begin with the axial local gauge defined by $A^+(x)=0$.
Notice that the non-local extension of this gauge is given by Eqn.~(\ref{CG-1}) provided the starting point is $x_0=-\infty$.
The differences between the local and non-local gauges can symbolically be demonstrated by the 
following trivial example. 
Consider two different (not arbitrary) vectors $A$ and $B$, they are different by construction. 
We now assume that these vectors 
have the same projection on the certain direction given by another vector $N$.
From the fact that the vectors have the same projections on the vector $N$,
it does not mean the equivalence of vectors $A$ and $B$, {\it i.e.}
\begin{eqnarray}
\Big\{ \text{if}\,\,\, (A\cdot N) = (B\cdot N)\Big\} \, \,\,\nRightarrow \Big\{ A=B \Big\}
\end{eqnarray} 
In this example, the different vectors $A$ and $B$ can be associated with two different contour (non-local) gauges,
\begin{eqnarray}
\label{A-B-1}
&&A\Leftrightarrow  \Big\{ {\bf g}(x | A)\Big |_{P(-\infty,x)}={\bf 1}\Big\}, \quad
\nonumber\\
&&
B\Leftrightarrow \Big\{ {\bf g}(x | A)\Big |_{P(+\infty,x)}={\bf 1}\Big\} ,
\end{eqnarray}
while the local axial gauge plays a role of the projections on $N$,
\begin{eqnarray}
\label{A-B-2}
\Big\{ (A\cdot N) = (B\cdot N)\Big\} \Leftrightarrow \Big\{ A^+(x)=0 \Big\}.
\end{eqnarray}
On the other hand, focusing on only the $N$ vector projection, there is no additional information 
to see from which vectors $A$ or $B$ the given projection has been performed.  

Coming back to the local axial gauge, we notice that
the local gauge suffers from the residual gauge transformations (which can lead to the
spurious pole uncertainties)
till the non-local contour gauge fixes all the gauge freedom in the finite space
(see for instance \cite{Anikin:2021oht}).
Indeed, consider the local axial gauge as an equation on the gauge function $\theta(x)$, {\it i.e.}
\begin{eqnarray}
\label{LGT-1}
A^{ +\, \theta}(x)= \omega(x)A^+ \omega^{-1}(x) + \frac{i}{g} \omega(x)\partial^+ \omega^{-1}(x)=0,
\end{eqnarray}
where $x=(x^-, \tilde x)$ with $\tilde x= (x^+, {\bf x}_\perp)$.
We can readily find a solution of this equation which takes the form of
the undetermined integration as
\begin{eqnarray}
\label{Sol-gauge-0}
&&\omega_0(x)=C \, \mathbb{P}\text{exp}
\Big\{  - ig\int dx^- A^+(x)\Big\},
\end{eqnarray}
or the form of determined integration as
\begin{eqnarray}
\label{Sol-gauge}
&&\omega_0(x^-, \tilde x)=C(\tilde x) \, \overline{\omega}(x^-, \tilde x),
\nonumber\\
&&
\overline{\omega}(x^-, \tilde x)= \mathbb{P}\text{exp}
\Big\{  - ig\int_{x_0^-}^{x^-} dz^- A^+(z^-, \tilde x)\Big\}.
\end{eqnarray}
Here, $x^-_0$ is fixed and $C(\tilde x)$ is an arbitrary function which does not depend on
$x^-$ and it is given by
\begin{eqnarray}
\label{C-norm}
\omega_0(x^-_0, \tilde x)=C(\tilde x).
\end{eqnarray}
The arbitrariness of $C$-function also reflects the fact that we here deal with an arbitrary fixed starting point $x_0$.
We now study the residual gauge freedom requiring both $A^{+,\, \theta}(x)=0$ and $A^{+}(x)=0$,
we then have
\begin{eqnarray}
\label{Sol-gauge-2}
&&\omega_0(x^-, \tilde x)\Big|_{\overline{\omega}=1}\equiv \omega^{\text{res}}(\tilde x)=
C(\tilde x)\equiv
e^{i\,\tilde\theta(\tilde x)}.
\end{eqnarray}
One can see that the function $C$ determines the residual gauge transforms.
This situation takes place in the local axial gauge defined by the only condition $A^+=0$
applied for (\ref{Sol-gauge}).

Now, we go over to the non-local gauge which in some sense gives more information 
on the gauge fixing.
The non-local contour gauge extends the local axial-type gauge
and demands that the full (curve-linear) integral in the exponential of Eqn.~(\ref{CG-1})
has to go to zero. We stress that  in contrast to the local gauge where the corresponding exponential disappears thanks to
that the integrand $A^+$ goes to zero. One can demonstrate it on the example of solutions (\ref{Sol-gauge-0}) or 
(\ref{Sol-gauge}).
Indeed, in the contour gauge
the residual gauge function $\tilde\theta(\tilde x)$ can be related to the path dependent functionals with  $A^-$ and $A^i_\perp$
which are also disappeared eliminating the gauge freedom and 
giving the physical gluon representation in the form of (\ref{A-cg}) (see \cite{Anikin:2016bor,Belitsky:2002sm} for details).
That is, if we restore the full path in the path dependent functional for a given process, we can get that
\begin{eqnarray}
\label{C-full}
&&C(\tilde x)=\tilde C(x^+_0, x^-_0, {\bf x}^\perp_0)
\nonumber\\
&&\times
\mathbb{P}\text{exp}
\Big\{ ig\int_{{\bf x}_0^\perp}^{{\bf x}^\perp} d \omega_\perp^i A^i_\perp(x^+_0, x^-_0, \omega_\perp)\Big\}
\nonumber\\
&&\times
\mathbb{P}\text{exp}
\Big\{ -ig\int_{x_0^+}^{ x^+} d \omega^+ A^-(\omega^+, x^-_0, {\bf x}_\perp)\Big\}.
\end{eqnarray}
Then, requiring the conditions as 
\begin{eqnarray}
\label{Con-cg-1}
&&A^-(\omega^+, x^-_0, {\bf x}_\perp)=0,
\\
&&
\label{Con-cg-2}
\int_{{\bf x}_0^\perp}^{{\bf x}^\perp} d \omega_\perp^i A^i_\perp(x^+_0, x^-_0, \omega_\perp)=0,
\end{eqnarray}
 we get that
\begin{eqnarray}
\label{C-full-2}
C(\tilde x)\Big|_{\text{c.g.}}=\tilde C(x^+_0, x^-_0, {\bf x}^\perp_0).
\end{eqnarray}
In the contour gauge, Eqn.~(\ref{C-full-2}) means that no the gauge freedom has left at all.
We emphasize that the condition of (\ref{Con-cg-1}) demands that the integrand is zero,
 while the condition of (\ref{Con-cg-2}) is imposed on the integration which leads to 
 the corresponding representation for the transverse gluon field, see (\ref{Ag1}),  (\ref{Ag2}) and 
 (\ref{A-cg}). Besides, 
the exact value of the fixed starting point $x_0$ depends on the process under
our consideration \cite{Belitsky:2002sm}.

The other feature of the path dependent functional is that it defines the path dependent gauge transformation
in the form of 
\begin{eqnarray}
\label{p-g}
&&A^{p.g.}_\mu(x) = 
{\bf g}^{-1}(x | A)\Big |_{P(x_0,x)}
A_\mu(x) {\bf g}(x | A)\Big |_{P(x_0,x)} + 
\nonumber\\
&&
\frac{i}{g}
{\bf g}^{-1}(x | A)\Big |_{P(x_0,x)} \partial_\mu {\bf g}(x | A)\Big |_{P(x_0,x)},
\end{eqnarray} 
where the starting point $x_0$ is now fixed. 
Hence, having calculated the derivative of ${\bf g}$ in (\ref{p-g}), we get that
(here, for the simplicity, $A_\mu(-\infty)=0$)
\begin{eqnarray}
\label{p-g-1}
&&A^{p.g.}_\mu(x)=
\int_{-\infty}^x dz_\alpha \frac{\partial z_\beta}{\partial x_\mu} 
\nonumber\\
&&
\times
{\bf g}^{-1}(z | A)\Big |_{P(-\infty,z)} 
\, G_{\alpha\beta} (z | A)\, 
{\bf g}^{-1}(z | A)\Big |_{P(-\infty,z)}, 
\end{eqnarray} 
where $x_0=-\infty$. 
If we are now focusing on the case of $x=(0^+, x^-, {\bf 0}_\perp)$, we readily obtain that
the gluon representation reads 
\begin{eqnarray}
\label{p-g-2}
A^{p.g.}_\mu(x^-)&=&
\int_{-\infty^-}^{x^-} dz^-  
[z^-; -\infty^-]_{A^+}^{-1} 
\nonumber\\
&&
\times
G^{+\mu} (z^- | A)\, [z^-; -\infty^-]_{A^+},
\end{eqnarray} 
where the traditional notations
\begin{eqnarray}
\label{WL-def}
[z_2 ;\, z_1]_{A}= {\bf g} (z_2 | A) \Big|_{P(z_1, z_2)}
\end{eqnarray}
have been introduced.
The representation given by (\ref{p-g-2}) is an important result for our further considerations.

So, in this section, we have demonstrated that the contour gauge gives a possibility to 
illuminate the unphysical gluon components, meanwhile the local axial gauge fixes the 
gauge partially leaving room for the residual gauge freedom.

\section{The advantage of the contour gauge use}
\label{Sec:V}

In the preceding sections, the different (mathematical) aspects of the contour gauge have been presented
in the general, even formal, manner.
In this section, we want to concentrate our attention
on the certain examples which, first, relate to the practical use of the contour gauge
and, second, demonstrate the preponderance of the non-local gauges compared to the local gauges.

As discussed in \cite{Anikin:2010wz,Anikin:2015xka,Anikin:2016bor}, the Drell-Yan-like processes
with the polarized hadrons give the unique example where
the contour gauge use shows the definite advantage from the practical point of view.
In particular, the contour gauge use allows to find the new contributions
to the Drell-Yan-like hadron tensor
which restore and ensure the gauge invariance of the corresponding hadron tensors \cite{Anikin:2010wz,Anikin:2015xka,Anikin:2016bor}.
It is important, however, to stress that the mentioned new contributions are invisible if we would work within the
frame in the local gauge.

In the similar manner, due to the contour gauge conception
the $\xi$-process of DVCS-amplitude which clarifies the gauge invariance
of the non-forward processes takes the closed form again  \cite{Anikin:2020ipg}.
From the practical point of view, it is instructive to consider the appearance of 
standard and non-standard diagrams contributing to the well-known deeply virtual Compton scattering (DVCS) amplitude
in the frame of the factorization procedure.
The gluons radiated from the internal quark of the hard subprocess
generate the standard diagrams, while the non-standard diagrams are formed by
the gluon radiations from the external quark of the hard subprocess (see \cite{Anikin:2020ipg} for details).

In the most cases, it is sufficient to exponentiate only the longitudinal components of gluon field, $A^-$ and $A^+$,
which are related to the unphysical degrees.  
Indeed, 
the standard diagram contributions give
the gauge invariant quark string operator which reads
\begin{eqnarray}
\label{OP1}
\bar\Psi^{(st)}(0^+,0^-,{\bf 0}_\perp | A^+)\,
\big\{ \gamma \big\}\,
\Psi^{(st)}(0^+,z^-,{\bf 0}_\perp | A^+),
\end{eqnarray}
where
\begin{eqnarray}
\label{Psi}
&&\bar\Psi^{(st)}(0^+,0^-,{\bf 0}_\perp | A^+)=
\nonumber\\
&&\bar\psi(0^+,0^-,{\bf 0}_\perp) [0^+,0^-,{\bf 0}_\perp; \, 0^+,+\infty^-,{\bf 0}_\perp]_{A^+},
\\
&&\Psi^{(st)}(0^+,z^-,{\bf 0}_\perp | A^+)=
\nonumber\\
&&
[0^+,+\infty^-,{\bf 0}_\perp; \, 0^+,z^-,{\bf 0}_\perp]_{A^+} \psi(0^+,z^-,{\bf 0}_\perp)
\end{eqnarray}
and $\big\{ \gamma \big\}$ stands for $\gamma$-matrices the exact form of which is now irrelevant.

The non-standard diagram contributions result in the string operator defined as
\begin{eqnarray}
\label{OP2}
\bar\Psi^{(non-st)}(0^+,0^-,{\bf 0}_\perp | A^-)
\, \big\{\gamma\big\}\,
\Psi^{(non-st)}(0^+,z^-,{\bf 0}_\perp | A^-),
\end{eqnarray}
where
\begin{eqnarray}
\label{Psi-non-st}
&&\bar\Psi^{(non-st)}(0^+,0^-,{\bf 0}_\perp | A^-)=
\nonumber\\
&&\bar\psi(0^+,0^-,{\bf 0}_\perp) [-\infty^+,0^-,{\bf 0}_\perp; \, 0^+,0^-,{\bf 0}_\perp]_{A^-},
\\
&&\Psi^{(non-st)}(0^+,z^-,{\bf 0}_\perp | A^-)=
\nonumber\\
&&[0^+,z^-,{\bf 0}_\perp; \, -\infty^+,z^-,{\bf 0}_\perp]_{A^-} \psi(0^+,z^-,{\bf 0}_\perp).
\end{eqnarray}
Hence, using the contour gauge conception,
we eliminate all the Wilson lines with the longitudinal gluon fields $A^+$ and $A^-$
demanding that
\begin{eqnarray}
\label{C-G-2}
&&[0^+,0^-,{\bf 0}_\perp; \, 0^+,+\infty^-,{\bf 0}_\perp]_{A^+} = {\bf 1},
\nonumber\\
&&[0^+,+\infty^-,{\bf 0}_\perp; \, 0^+,z^-,{\bf 0}_\perp]_{A^+} = {\bf 1}
\end{eqnarray}
and
\begin{eqnarray}
\label{C-G-3}
&&[-\infty^+,0^-,{\bf 0}_\perp; \, 0^+,0^-,{\bf 0}_\perp]_{A^-} = {\bf 1},
\nonumber\\
&&[0^+,z^-,{\bf 0}_\perp; \, -\infty^+,z^-,{\bf 0}_\perp]_{A^-} = {\bf 1}.
\end{eqnarray}
Eqns.~(\ref{C-G-2}) and (\ref{C-G-3}) give rise to the local gauge conditions given by $A^+=0$
and $A^-=0$.

With respect to the Wilson line with the transverse gluons $A^i_\perp$, we remind
that we work here within
the factorization procedure applied for DVCS-amplitude.
In this case, the Wilson lines with the transverse gluon fields
are considered in the form of an expansion due to the fixed twist-order,
and the transverse gluons correspond to the physical configurations of L-system.

Thus, the DVCS process gives us the example how the unphysical gluon degrees can be 
illuminated from the consideration with the help of contour gauge.

We are now going over to the Drell-Yan (DY) process with one transversely polarized hadron \cite{Anikin:2010wz}
\begin{eqnarray}
\label{DY}
N^{(\uparrow\downarrow)}(p_1) + N(p_2)\Rightarrow
\ell(l_1) + \bar\ell(l_2) + X(P_X)\,,
\end{eqnarray}
where the virtual photon producing the lepton pair ($l_1+l_2=q$) has a large offshelness, {\it i.e.}
$q^2=Q^2\to \infty$,
while all the transverse momenta are small and integrated out in the corresponding cross-sections $d\sigma$.
Here, the contour gauge use results in the gauge invariant hadron tensor and provides the 
new contributions to single spin asymmetries.
Having considered this hadron tensor in 
the asymptotical regime associated with the very large $Q^2$, 
the factorization theorem can be applied for the given hadron tensor as well as for the DVCS process.
As a result, the DY hadron tensor takes a form of convolution as
\begin{eqnarray}
\label{Fac-DY}
&&Hadron\,\, tensor =
\nonumber\\
&&
\{Hard\,\, part\,\, (pQCD)\} \otimes
\{Soft \,\,part\,\, (npQCD) \}\,,
\end{eqnarray}
where both the hard and soft parts should be independent of each other and are in agreement with
the ultraviolet and infrared renormalizations.
Moreover, the relevant single spin asymmetries (SSAs), which is a subject of experimental studies,
can be presented as
\begin{eqnarray}
d\sigma(N^{(\uparrow)}) - d\sigma(N^{(\downarrow)})
\sim {\cal L}_{\mu\nu}\, H_{\mu\nu}\, ,
\end{eqnarray}
where ${\cal L}_{\mu\nu}$ and $H_{\mu\nu}$ are the lepton and
hadron tensors, respectively.
The hadron tensor $H_{\mu\nu}$ includes the
the polarized hadron matrix element which takes a form of
\begin{eqnarray}
\label{parVecDY}
&&\langle p_1, S^\perp | \bar\psi(\lambda_1 \tilde n)\, \gamma^+ \,
g A^{\alpha}_\perp(\lambda_2\tilde n) \,\psi(0)
|p_1, S^\perp \rangle
\stackrel{{\cal F}}{=}
\nonumber\\
&&
i\varepsilon^{\alpha + S^\perp -} (p_1p_2)
\, B(x_1,x_2),
\end{eqnarray}
where $\stackrel{{\cal F}}{=}$ stands for the Fourier transform between
the coordinate space, formed by positions $\lambda_i \tilde n$, and
the momentum space, realized by fractions $x_i$;
the light-cone vector $\tilde n$  is a dimensionful analog of $n$.
In Eqn.~(\ref{parVecDY}), the parametrizing function $B$ describes the corresponding parton distribution.

\subsection{The case of local gauge}

In the studies, see for example \cite{Boer:2003cm,BQ,Teryaev,Boer}, where the local light-cone  gauge $A^+=0$
has been used,
$B$-function of Eqn.~(\ref{parVecDY}) is given by a purely real function. That is, we have
\begin{eqnarray}
\label{g-pole-B}
B(x_1,x_2)= \frac{{\cal P}}{x_1-x_2} T(x_1,x_2)\,
\end{eqnarray}
where the function $T(x_1,x_2)\in\Re\text{e}$
parametrizes the corresponding projection of $\langle \bar\psi\, G_{\alpha\beta}\,\psi \rangle$ and
obeys $T(x,x)\not = 0$.

In Eqn.~(\ref{g-pole-B}), the pole at $x_1=x_2$ is treated as a principle value and it obviously leads to $B(x_1,x_2)\in\Re\text{e}$.
Indeed, within the local gauge $A^+=0$,
the statement on that $B$ is the real function stems actually from
the ambiguity in the solutions of the trivial differential equation, which is equivalent to the definition of $G_{\mu\nu}$,
\begin{eqnarray}
\label{DiffEqnsG}
\partial^+\, A^\alpha_\perp=G^{+\,\alpha}_\perp\,.
\end{eqnarray}
The formal resolving of Eqn.~(\ref{DiffEqnsG}) leads to two representations written as
\begin{eqnarray}
\label{Ag1}
&&A^\mu_{(1)}(z) =
\int_{-\infty}^{z} d\omega^-
G^{+\mu} (\omega^-)
+ A^\mu(-\infty),
\\
\label{Ag2}
&&A^\mu_{(2)}(z) =
- \int_{z}^{\infty} d\omega^-
G^{+\mu} (\omega^-)
+ A^\mu(\infty).
\end{eqnarray}

We stress that within the approaches backed on the local axial gauge use, there are no evidences to think that
Eqns.~(\ref{Ag1}) and (\ref{Ag2}) are not equivalent each other.
That is, the local gauge $A^+=0$ inevitably leads to the following logical scheme (see \cite{Anikin:2015xka} for details)
\begin{eqnarray}
\label{Loc-g-equiv}
\boxed{A^\mu_{(1)} \stackrel{loc.g.}{=}A^\mu_{(2)}} \rightarrow
\boxed{B(x_1,x_2)={\cal P}\frac{T(x_1,x_2)}{x_1-x_2} \in\Re\text{e}}.
\end{eqnarray}
This equation demonstrates that the equivalence of Eqns.~(\ref{Ag1}) and (\ref{Ag2})
causes the representation of $B$-function as in Eqn.~(\ref{g-pole-B}).
The discussion on the boundary configurations can be found in \cite{Anikin:2015xka}.

Regarding the DY process, the physical consequences of the use of $B$ presented by Eqn.~(\ref{g-pole-B}) are
the problem with the photon gauge invariance of DY-like hadron tensors and the losing of significant contributions
to SSAs \cite{Anikin:2010wz,Anikin:2015xka}.
Besides, based on the local gauge $A^+=0$, the representation of gluon field as a linear
combination of Eqns.~(\ref{Ag1}) and (\ref{Ag2}) has been used in the different studies, see
\cite{Belitsky:2005qn,Hatta:2011zs,Lorce:2013gxa}.

\subsection{The case of non-local gauge}

In contrast to the local gauge $A^+=0$, as discussed in Sec.~\ref{Sec:III}, we can infer that
the path dependent non-local gauge (see Eqn.~(\ref{CG-1}))
fixes unambiguously the representation of gluon field which is given by either Eqn.~(\ref{Ag1}) or Eqn.~(\ref{Ag2}).
Indeed, fixing the path $P(x_0,x)$, a solution of Eqn.~(\ref{CG-1}) takes the form of
(see \cite{Anikin:2010wz,Anikin:2015xka,Durand:1979sw} for details)
\begin{eqnarray}
\label{A-cg}
A^{c.g.}_\mu(x)=
\int_{P(x_0,x)} dz_\alpha \frac{\partial z_\beta}{\partial x_\mu}\,
G_{\alpha\beta}(z| A),
\end{eqnarray}
where the boundary configuration $A^{c.g.}_\mu(x_0)$ has assumed to be zero.
By direct calculation, we can show that the non-local gauge ${\bf g} (P(- \infty, x) | A)=1$ leads to the gluon filed
representation of Eqn.~(\ref{Ag1}),
while the non-local gauge ${\bf g} (P(x, + \infty ) | A)=1$ corresponds to the gluon field representation of Eqn.~(\ref{Ag2}).
Moreover, we can readily check that \cite{Anikin:2010wz,Anikin:2015xka}
\begin{eqnarray}
\label{2-cg}
\text{Eq.~(\ref{Ag1})}\rightarrow \boxed{B_{+}(x_1,x_2)} \not=
\boxed{ B_{-}(x_1,x_2)} \leftarrow \text{Eq.~(\ref{Ag2})},
 \end{eqnarray}
where
\begin{eqnarray}
\label{B-plus-minus}
&&
B_{+}(x_1,x_2)=\frac{T(x_1,x_2)}{x_1-x_2 + i\epsilon}\in \mathbb{C},
\\
&&
B_{-}(x_1,x_2)=
\frac{T(x_1,x_2)}{x_1-x_2 - i\epsilon}\in \mathbb{C}.
\end{eqnarray}

Notice that both the non-local contour gauges, {\it i.e.} ${\bf g} (P(- \infty, x) | A)={\bf 1}$ and 
${\bf g} (P(x, + \infty ) | A)={\bf 1}$,
can be projected into
the same local gauge $A^+=0$. As mentioned, the projection given by $A^+=0$ does not give a possibility
to understand which of the non-local gauges generates the local gauge.

To conclude, we can state that, considering DY-like processes,
the corresponding non-local gauge gives rise to the correct representation
of $B$-functions, see Eqn.~(\ref{B-plus-minus}), which has the non-zero imaginary part.
This enables us to find the new significant contributions to the hadron tensors that ensure ultimately
the gauge invariance  \cite{Anikin:2010wz,Anikin:2015xka}. In a similar manner, with the help of
Eqn.~(\ref{B-plus-minus}) we can fix the prescriptions for
the spurious singularities in the gluon propagators \cite{Anikin:2016bor}.

\section{Conclusions}

In the paper, we have made public the important subtleties based on the mathematical   
technique adjusted to the physical language.  
We have presented the important explanations and analysis hidden in the preceding publications
which should help to clarify 
the main advantages of the use of non-local contour gauges.
To this goal, the combination of the Hamilton and Lagrangian approaches to the
gauge theory has been exploited in our consideration.
Since the contour gauge is mainly backed on the geometrical interpretation of gluon fields as a connection on the
principle fiber bundle, we have provided the illustrative demonstration of geometry of the contour gauge.
In this connection, the Hamiltonian formalism is supposed to be more convenient for understanding
the subtleties of contour gauges. Indeed, the Lagrange factor $\lambda_a$ fixation has a direct treatment
in terms of the orbit group element which has been uniquely chosen by the corresponding plan transecting
the principle fiber bundle, see Fig.~\ref{Fig-S-2}.
While, as shown, the Lagrangian formalism is very well designed for the practical uses to the eliminate
the unphysical gluon degree of freedom from the corresponding amplitudes.

Also, we have reminded the details of that studies where the local axial-type gauge can lead to the certain ambiguities
in the gluon field representation. These ambiguities may finally produce incorrect results.
Meanwhile, as demonstrated in the paper, the non-local contour gauge can fix this kind of problems and, for example,
can provide not only the correct (gauge invariant) final result but also find the new contributions to
the hadron tensors of DY-like processes \cite{Anikin:2010wz,Anikin:2015xka}.
Thus, the use of contour gauge conception gives a possibility
(i) to find a solution of the gauge-invariance problem, discovered in the Drell-Yan hadron tensor, by the correct description
of the gluon pole that is appearing in the corresponding parton distributions;
(ii) to discover the new sizeable contributions to the single-spin asymmetries which are under intensive experimental studies.

In the context of the contour gauge use, the recent progress is mainly related to
the studies of the so-called gluon pole contributions to the
Drell-Yan-like processes \cite{Anikin:2010wz,Anikin:2015xka}. However, the practical profit
of the non-local contour gauge is not limited by the study of gluon poles which manifest
in the different hard processes. With the help of non-local gauges, we plan to adopt the
method based on the geometric quantization \cite{Nair:2016ufy} to the investigation of
different asymptotical (hard) regimes in QFT.

{\bf Acknowledgements.}
We thank Y.~Hatta, C.~Lorce, A.V.~Pi\-mikov, L.~Szymanowski, O.V.~Teryaev, A.S.~Zhevlakov 
and colleagues from the Theoretical Physics Division of NCBJ (Warsaw)
for useful and stimulating discussions.
This work is supported by the Ulam Program of NAWA No.
PPN/ULM/2020/1/00019.


\end{document}